**Effect of feathers on drag in plunge-diving birds**

Florent Debenedetti[1], Sunghwan Jung[2]

[1]Département de Mécanique, Ecole Polytechnique, Palaiseau, France

[2]Department of Biological and Environmental Engineering, Cornell University, Ithaca, New York, USA

**Correspondence**

Shungwan Jung

Department of Biological and Environmental Engineering,

Cornell University,

111 Wing Dr.

Ithaca, NY 14853, USA

sj737@cornell.edu

**Graphical Abstract**

This study examines how feathers influence the hydrodynamic drag in diving birds. It reveals that the drag coefficient during water entry is doubled by feathers, likely due to the roughness of feather structures. Despite the feathers' ability to hold and release bubbles, there is no reduction in drag. These findings could guide the development of bio-inspired aquatic vehicles.

## Abstract

This study explores the impact of feathers on the hydrodynamic drag experienced by diving birds, which is critical to their foraging efficiency and survival. Employing a novel experimental approach, we analyzed the kinematics of both feathered and non-feathered projectiles during their transition from air to water using high-speed imaging and an onboard accelerometer. The drag coefficients were determined through two methods: a direct calculation from the acceleration data and a theoretical approach fitted to the observed velocity profiles. Our results indicate that feathers significantly increase the drag force during water entry, with feathered projectiles exhibiting approximately double the drag coefficient of their smooth counterparts. These findings provide new insights into the role of avian feather morphology in diving mechanics and have potential implications for the design of bio-inspired aquatic vehicles in engineering. The study also discusses the biological implications of increased drag due to feathers and suggests that factors such as body shape might play a more critical role in the diving capabilities of birds than previously understood.

**KEYWORDS** diving, drag, diving birds, feather

# INTRODUCTION

In nature, some animals exhibit diving behaviors with diverse body shapes and surface coverings that have evolved to optimize their interaction with water [1, 2]. When an animal dives the shape of the diving front significantly influences the forces encountered; a blunt front tends to increase drag and impact force, whereas a sharp, narrow front can reduce these forces [3, 4]. Over evolutionary timescales these pressures have the potential to alter animal morphology, driving adaptations that minimize resistance and impact during diving [5].

Avian feathers play a critical role in both the aerodynamics and hydrodynamics of birds. Notably, observations of penguins show that they create a trail of air bubbles when leaping from water to ice shelves. This phenomenon is thought to involve air trapped under feathers that is released to lower drag during the jump [6]. Micro-bubbles, which attach to and are emitted from the hydrophobic feathers [7], indicate the free-shear boundary condition on the body, thereby reducing drag [8–14]. Additionally, plunge-diving birds have been observed to release bubbles from the air trapped in their underlying feathers (see Fig. 1a ) as they descend while plunging and rapidly swimming ~~to the surface as shown in Fig. 1a~~ [3]. Despite some studies demonstrating a relationship between drag and the feathers of surface-diving birds [15, 16], the effect of the bubble release of plunge-diving birds on drag remains unexplored. Additionally, feathers are known to spread impact force through elastic couplings, providing further functional advantages [17]. Previous research has contrasted drag on feathered bodies with that on smooth ones in a fully immersed tank [15, 18, 19]. These studies consistently found that feathered bodies experience higher drag than their smooth counterparts, attributable to feather roughness and fluttering. This aligns with other research indicating that feathers enhance lift force, stabilize flight, and temporarily amplify drag during landing [20, 21].

Hydrodynamic resistance on a projectile is closely linked to its overall shape (form drag), with surface properties also playing a significant role (skin friction) [22]. Given that all birds are covered in feathers, these features potentially influence skin friction during plunge-diving [3]. This paper considers two potential effects of feathers on hydrodynamics. Firstly, birds can trap air beneath their feathers when diving, modifying the surface characteristics. Air bubbles have been shown to decrease drag by reducing shear resistance on the body surface [23–25]. Secondly, the hydrophobic nature of feathers [26] is another crucial aspect. Hydrophobic surfaces are known to lessen drag by creating a slip condition on the surface [27, 28]. Prior studies suggest that these features could help reduce skin friction on feathered surfaces.

The above summary highlights a paradox in avian hydrodynamics: while both air bubbles and hydrophobic surfaces associated with feathers can reduce drag, the inherent roughness and fluttering motions of feathers conversely increase it. Furthermore, accurately measuring the drag coefficient on live or deceased birds presents significant experimental challenges [29]. This difficulty arises from the disturbance of flow patterns caused by harnesses or struts attached to the birds during testing, complicating the collection of reliable data.

In this present study, we will investigate how feathers affect drag during plunge diving.  Our experiment was designed for a projectile to plunge dive into water from air.  The projectile was equipped with an accelerometer and had exteriors covered with feathers or without feathers and smooth surface.  The drag coefficient was evaluated from the measured acceleration and velocity, and the biological and physical implications of feathered objects are discussed.

## MATERIALS AND METHODS

### Experimental setup

The experimental setup featured a drop tower with a water tank, a projectile equipped with an accelerometer, and a light source, as depicted in Figure 1b.  The tank, made of one-inch thick clear acrylic for optical clarity above and below the water surface, measured 54×54×135 cm internally.  An adjustable projectile releaser was mounted on a fixed frame allowing for the control of the impact speed by varying the drop height.  This releaser [30] comprised a manually operable camera diaphragm, supplemented by small acrylic plates for guiding the projectile in a straight path.  The projectile was placed on this device and released by gradually opening the diaphragm until the aperture was larger than the projectile, allowing it to drop solely under gravity.  Upon release, the projectile accelerated downwards, impacting and entering the water with its descent observable down to the tank's bottom.

A plastic net was placed at the tank's bottom to gently cushion the projectile upon impact and prevent damage. This choice proved less disruptive to the water flow compared to foam, as water could easily pass through the net and minimize side effects. The setup's design ensured a consistent, repeatable release mechanism. For precise measurement of the projectiles' kinematics as they crossed the water surface, the experiment employed both a high-speed camera (Photron Fastcam Mini UX100) and an onboard accelerometer. This combination enabled detailed observation and data collection regarding the projectile's motion and behavior upon entering the water.

## Projectiles

We investigated the drag on a free-falling projectile adorned with bird feathers. Our experiments involved dropping two types of projectiles, identical in size but differing in surface texture: one with feathers and the other with a smooth body, as illustrated in Fig. 1c. The feather patch used in the experiments was obtained from a bird that had naturally died. This bird was collected by a staff member from the North Carolina Museum of Natural History. The dissection procedures were carried out at the Smithsonian Museum of Natural History. A careful dissection was performed to separate the skin (including feathers) from the bones and muscles around the bird's neck. Then, the feather patch was superglued onto the projectile, completing its assembly. The feather density of the same bird feather has been characterized in Ref. 17. It has been observed that the calami of the feathers are arranged in a hexagonal pattern, with an approximate distance of 5.3 mm between each calamus.

The projectile was composed of three parts. Its lower section was a $20°$ cone, mirroring the average angle of a bird's skull. The middle section was a cylinder, to which either bird feathers or a smooth surface were attached. We tested two different top shapes. Initially, we used acrylic plastic caps, laser cut to the precise diameter (see the left panel of Fig. 1c). However, this resulted in strong pinch-off dynamics in a trailing air column and induced ripple dynamics. Consequently, the acceleration/force signal varied significantly, making it difficult to discern the effects of surface texture. Next, we tried a profiled shape cap (see the right panel of Fig. 1c). With this design, we observed minimal force oscillation due to the pinch-off. Therefore, the top part was designed for stability and integration with the release mechanism.

All three components were constructed using a 3D printer (Makerbot Replicator 2X, utilizing ABS filaments). The projectile had a radius of 3.5 cm, resulting in a cross-sectional area of 38.5 cm$^2$. This design enabled a controlled comparison of the drag effects between feathered and smooth surfaces.

**On-board accelerometer**

The onboard accelerometer, a crucial component of our experimental apparatus, offered a high degree of precision in measuring acceleration, a task that often presents challenges when using high-speed images alone. Our customized setup included a 10-bit accelerometer (ADXL 335; ±3g range), an SD card writer (Adafruit Co.; part number 254), and an Arduino MICRO, all compactly configured into a unit with dimensions of 2.9×3.8×4.5 cm$^3$ (See Fig. 2). This compact assembly allowed for seamless integration into the interior of the projectiles without affecting their flight dynamics. With a notable sampling rate of approximately 750 Hz and a fine acceleration resolution of 0.09 m/s$^2$, our system was able to capture the subtle nuances of projectile motion with high fidelity. The data obtained from the accelerometer was rigorously compared to that acquired from a high-speed camera to ensure accuracy.

Adjustments were made to the accelerometer data to account for the angular position of the projectiles (as observed in the high-speed camera footage) since any tilt in the object's orientation could slightly skew the readings.

The study also uncovered issues related to the sensitivity of the accelerometer. The device's resolution was pivotal when considering the variations in acceleration, particularly in scenarios involving feathered projectiles where air bubbles could cause disturbances, leading to increased standard deviation in the calculated drag coefficient. With a resolution of 0.09 m/s$^2$ and an associated uncertainty of approximately 0.03 m/s$^2$ in the drag coefficient, improving sensor sensitivity could enhance measurement precision. For instance, utilizing an accelerometer with a ±2 g range and 14-bit resolution could significantly reduce data uncertainty, although this was not tested due to the limitations of our current Arduino micro setup, which supports only 10-bit inputs. A 14-bit accelerometer would theoretically offer a 24-fold improvement in resolution.

Similarly, the high-speed camera data were also refined. The direct capture of positional information by the camera introduced noise when differentiating to obtain velocity; hence, a smoothing process was implemented to produce cleaner data. The results from both the high-speed camera and the onboard accelerometer, as shown in Figure 3, displayed remarkable consistency and validated the precision of our accelerometer in capturing the velocity of the projectiles. This cross-verification process was critical to ensuring that any discrepancies between the two measurement methods remained inconsequential, thereby reinforcing the reliability of our data acquisition approach.

**Equation of motion**

When a projectile is fully submerged, it is subject to three predominant forces: gravity, buoyancy, and drag. The equation governing its underwater motion encapsulates the interplay of these forces and can be articulated as follows:

$$(M + M_{\text{add}})\, \frac{dV}{dt} = Mg - F_B - \frac{1}{2}\rho C_d\, V^2 S\,.  \tag{1}$$

Here, $M$ is the mass of the projectile, $M_{\text{add}}$ is the added mass, $V$ is the descending velocity, $g$ is the gravitational acceleration, $F_B$ is the buoyancy force, $\rho$ is the fluid density, and $S$ is the cross-sectional area of the projectile. The concept of added mass, $M_{\text{add}}$, is important in fluid dynamics as it characterizes the additional inertia resulting from the fluid that must be accelerated along with the object [22]. In this context, $M_{\text{add}}$ is estimated to be half the buoyancy force divided by gravitational acceleration, given by the relation $M_{\text{add}} = F_B/2g$.

The dynamic behavior of the projectile under water is discerned in two distinct regimes: jerk and steady regimes. The jerk regime is defined as a rapid change in acceleration. During the jerk regime (shaded in red in Fig. 5a), the forces of gravity and buoyancy are predominant, with drag playing an inconsequential role due to the relatively trivial velocity. Simplifying Equation (1) under these conditions yields the following relation:

$$(M + M_{\text{add}})\, \frac{dV}{dt} = Mg - F_B  \tag{2}$$

which upon integration, assuming initial conditions where velocity is zero at time zero, provides a linear velocity profile:

$$V(t) \ = \ A_{\text{body}} \, t \ = \ \left( \frac{Mg - F_B}{M + M_{\text{add}}} \right) t \, , \tag{3}$$

with $A_{\text{body}}$ delineating the net acceleration exerted by body forces. This expression for velocity as a function of time is instrumental for inferring the buoyancy force acting on the projectile—crucial for subsequent analysis and discussions. Transitioning to the steady regime (shaded in blue in Fig. 5a), the projectile attains a velocity that remains constant, effectively nullifying both acceleration and deceleration. The equation governing this state of motion asserts that the gravitational force minus the buoyancy force is balanced by the drag force

$$Mg \, - \, F_B \, = \, \frac{1}{2} \, \rho \, C_d \, V^2 \, S \qquad\qquad (4)$$

The steady velocity becomes

$$V_{\text{steady}} = \sqrt{\frac{2(Mg - F_B)}{\rho C_d S}} \qquad\qquad (5)$$

The interval required to shift from the jerk regime to the steady regime, termed the transition time, $T_{\text{transition}}$, is  deduced by equating the linear velocity from the jerk regime to the steady velocity. This gives:

$$T_{\text{transition}} = \frac{V_{\text{steady}}}{A_{\text{body}}} \qquad (6)$$

Employing typical values from the experiment ($C_d \sim 0.3$, $M \sim 0.9$ kg, $M_{\text{add}} \sim 0.4$ kg, and $F_B \sim 8$ N), we can infer that the steady velocity is in the vicinity of 1.5 m/s and the body acceleration is roughly 0.6 m/s$^2$. These numbers imply that the projectile would necessitate a travel distance exceeding 3 meters underwater to attain this steady velocity, considering a transition time in the neighborhood of 3 seconds.

**RESULTS**

In this section, we aimed to quantify the influence of feathers on drag. Specifically, to isolate the shear-drag effect attributable to feathers alone, we had engineered two distinct types of projectiles—one adorned with feathers and the other with a smooth surface— while ensuring that all other components remained identical in terms of materials and structural design.

**Observations**

The experiment involved dropping both smooth and feathered projectiles into a water tank. The precise moment of release and subsequent water entry were recorded, revealing initial stationary conditions followed by gravitational acceleration and a marked deceleration upon water contact. Figure 4 shows a sequence of high-speed camera snapshots that depict the projectiles at various stages of their drop, spaced 50 ms apart. This visual observation likely illustrates the distinct trajectories of the smooth versus feathered projectiles as they pierce the water surface.

Figure 5 detailed the recorded acceleration and velocity data from the experiment. In the plot, the red and black lines represent the smooth and feathered projectiles, respectively. Prior to point A (release moment) the flat line signifies that the projectiles are at rest. As the diaphragm releases the projectiles, a sharp increase in acceleration to match the force of gravity is observed. This is evident in the steep ascent of the graph lines up to the point of water entry at B, where a sudden deceleration occurs. Figure 5 illustrates the divergence in acceleration and velocity profiles post water entry, highlighting the variance in drag effects due to the different surface textures.

The post-entry behavior showed the feathered projectile experiencing a more complex interaction with the water, as indicated by the strong descent in acceleration, possibly due to the release of air bubbles or the interaction of feathers with the fluid, contrasting with the deceleration profile of the non-feathered projectile. This difference underscores the role of feathers in modifying the drag forces acting on the body during water entry.

**Buoyancy and drag forces**

In addition to the plunge-diving projectile experiment, a separate experiment was essential to ascertain the volume of air trapped beneath the feathers. Without this, our single acceleration measurement would have been insufficient to simultaneously determine both these parameters. To ascertain the buoyancy force, experiments were conducted wherein a projectile was dropped from an already submerged position. This setup ensures that the projectile accelerates under the influence of its own weight while maintaining minimal velocity. Consequently, this experiment showed an almost constant acceleration, providing a controlled environment to accurately gauge the effects of buoyancy only.

In Figure 6, the depicted vertical velocities exhibit a linear relationship with time where drag forces are yet to play a significant role due to a low speed. The acceleration, which is the gradient of the velocity–time graph, is contingent upon the mass of the projectile and the buoyancy force, consistent with the relationship described by Equation (3). With known projectile masses, the buoyancy force can be deduced. The graph shows that the smooth projectile undergoes greater acceleration compared to the feathered one, which is reflected in the steeper slope of its velocity–time curve.

Empirical measurements yield buoyancy forces of $8.14 \pm 0.08$ N for the smooth projectiles across nine trials, and $7.66 \pm 0.08$ N for the feathered projectiles over five trials. The marginally reduced buoyancy force observed in the feathered projectiles can be ascribed to the structural properties of feathers; they are porous and composed of slender structures like rachis and vanes, which may affect the displacement of water and consequently the buoyant force experienced by the projectile.

**Fit with an exact solution**

This approach to calculating the drag coefficient involved a curve fitting process that aligns the velocity profiles of the projectiles with the theoretical model. Assuming constant values for the projectile mass, buoyancy force, and drag coefficient over the duration of the experiment, we refer to the fundamental equation of motion, Eq. (1), to derive an exact solution for velocity as a function of time:

$$V(t) = V_{\text{steady}} \coth\left[\frac{t - t_0}{T_{\text{transition}}} + \text{arccoth}\left(\frac{V(t = t_0)}{V_{\text{steady}}}\right)\right]$$

(7)

This solution stipulates that if the initial velocity $V(t = t_0)$ is less than the steady-state velocity $V_{\text{steady}}$, the hyperbolic cotangent function in the equation is substituted with a hyperbolic tangent function. The equation presents two variables to be determined: the drag coefficient and the buoyancy force.

The curve fitting was executed using Matlab, adjusting the theoretical curve to match the experimental velocity data. The fitting process targets the maximization of the correlation coefficient, thereby deriving the most probable value for the drag coefficient. The fitting is consistently performed using a time window of 75 ms to align with the temporal scope of previous measurements. As shown in Figure 7, the experiment showed that the presence of feathers results in a higher drag coefficient when compared to the smooth surface. This finding consistently supports the notion that feathers influence the drag characteristics of the projectile.

## DISCUSSION AND CONCLUSION

In this study, we examined the effect of feathers on the drag experienced by projectiles using an onboard accelerometer to measure dynamic responses. The buoyancy force, inherently dependent on the fluid volume displaced by an object, is influenced by the porous nature of feathers and introducing complexity into the measurement process.

We determined that the drag coefficient for feathered projectiles was approximately double that of smooth projectiles, according to two distinct methodologies. Notably, this substantial difference in drag coefficients is not predominantly governed by buoyancy force variations. When a standardized buoyancy force ($F_B = 8.14$ N) is applied to both projectile types, the resulting drag coefficient exhibits a marginal 5% discrepancy, suggesting other influential factors are at play.

The findings indicated that feathers contribute to increased drag, potentially limiting a bird's ability to dive deeply. For deeper dives, birds would need to increase their entry velocity into the water, which inherently carries a risk of injury due to higher impact forces. The ability of birds to dive deeper than humans is therefore likely attributed to factors other than feathers, such as body shape, which appears to play a crucial role.

During the experiments, we had utilized two projectile types—one with feathers and one smooth—both sharing a similar shape when dry. However, upon water entry, the expulsion of air altered the shape of the feathered projectile, adding another layer of complexity to the analysis of drag forces and highlighting the dynamic nature of feathered surfaces in fluid environments.

Although our study showed the effect of bird feathers on drag, the detailed effects of each type of feather remain elusive. We believe that downy feathers could retain numerous bubbles, while contour feathers could release bubbles through gaps. Although our current experimental setup does not allow for modification of feather configuration or types, future studies could explore these aspects in more detail. Such investigations could provide valuable insights into the design of aquatic equipment and vehicles, potentially leading to significant advancements in reducing hydrodynamic drag. Furthermore, understanding the role of different feather types in drag reduction could also contribute to our knowledge of evolution and adaptation strategies of plunge-diving birds.

The current study utilized a feather-skin patch from a salvaged bird. Initially, the patch formed a cohesive shape with interconnected hooklets on barbules. However, over several runs of experiments, the feathers became floppy and lost the connections between barbules, indicating both physical and chemical degradation. In this study, we performed 15 trials, with about an extra 50 trials to optimize our setup and devices with the same feather patch. In the future, if we can obtain enough fresh feather-skin patches, we might be able to achieve results with less degradation in feather samples

## ACKNOWLEDGMENTS

The authors thank Dr. Brian Chang for his initial contribution to this project, Dr. Lorian Straker, Dr. Carla Dove for feather samples, and Dr. Christophe Clanet for the support of Mr. Florent Bedenedetti. This work was supported by the National Science Foundation Grant No. CBET-2002714.

## AUTHOR CONTRIBUTIONS

S.J. conceived the idea; F.D. and S.J. designed and built the experiment, and collected and analyzed the data. F.D. wrote the manuscript, and S.J. revised and edited the manuscript.

**COMPETING INTERESTS**

The authors have no competing interests.

**FIGURE LEGENDS**

**FIGURE 1** (a) Diving of a gannet (*Morus bassanus*), (b) the experimental setup, and (c) projectile variants used for drop tests. (b) Illustrates the drop tower mechanism with the projectile positioned at the release point. (c) Showcases the two categories of projectiles used: smooth and feathered projectiles. The first pair is equipped with flat acrylic caps and was utilized in preliminary experiments. The right pair with a very elongated profiled cap was used in the final experiments to calculate the drag coefficient ($C_d$).

**FIGURE 2** Photograph of the onboard accelerometer assembly, consisting of a microcontroller and an ADXL335 accelerometer module alongside a standard 9V battery. The scale bar represents 1 cm.

**FIGURE 3** Velocity–time graph comparing measurements from the onboard accelerometer (solid line) and high-speed camera (circular dots). The red line shows a linear velocity increase due to gravity only.

**FIGURE 4** Sequential high-speed camera images documenting the water entry dynamics of the two different projectile types. (a) The top row depicts the motion of a smooth projectile, while (b) the bottom row shows a feathered projectile. Each column represents a time step of 50

ms starting from the moment of water entry at $t = 0$ ms and ending at $t = 250$ ms. The images capture the distinct interaction patterns of each projectile with the water, illustrating the variations in splash, cavitation, and water flow caused by the different surface textures.

**FIGURE 5** Comparative analysis of acceleration (a) and velocity (b) for smooth and feathered projectiles during a drop from a height of 105 cm. The black and red lines represent the feathered and smooth projectiles, respectively. In graph (a), point A marks the release, and point B indicates water entry, where a notable difference in acceleration between the two types is observed. The pinch-off event is marked for the feathered projectile, indicating a phase of separation within the fluid. Graph (b) displays the corresponding velocity profiles, highlighting the differences in deceleration post water entry.

**FIGURE 6** Velocity versus time curves for projectiles submerged in water, showcasing the difference in acceleration between smooth and feathered designs. The solid line represents the smooth projectile, characterized by an acceleration of 1.34 m/s$^2$ and a buoyancy force ($F_B$) of 8.09 N. The dashed line corresponds to the feathered projectile, which exhibits a lower acceleration of 1.16 m/s$^2$ and a buoyancy force of 7.83 N, indicating the effect of the feather's texture on the hydrodynamic behavior.

**FIGURE 7** Drag coefficient ($C_d$) as a function of Reynolds number for two types of projectiles. The black squares represent the data for feathered projectiles, while the red circles indicate the data for smooth projectiles. Error bars signify the standard deviation of measurements across multiple trials illustrating the variation in $C_d$ at different Reynolds numbers within the tested range. We had five trials with feathered projectiles and 10 trials with smooth projectiles.

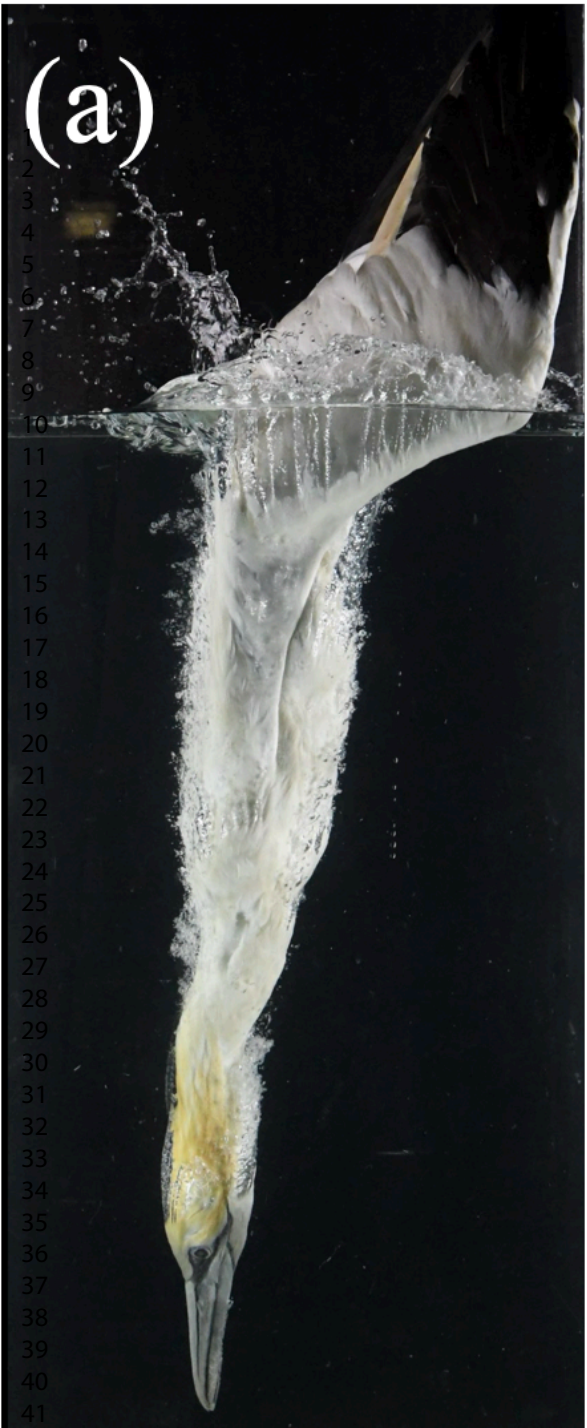
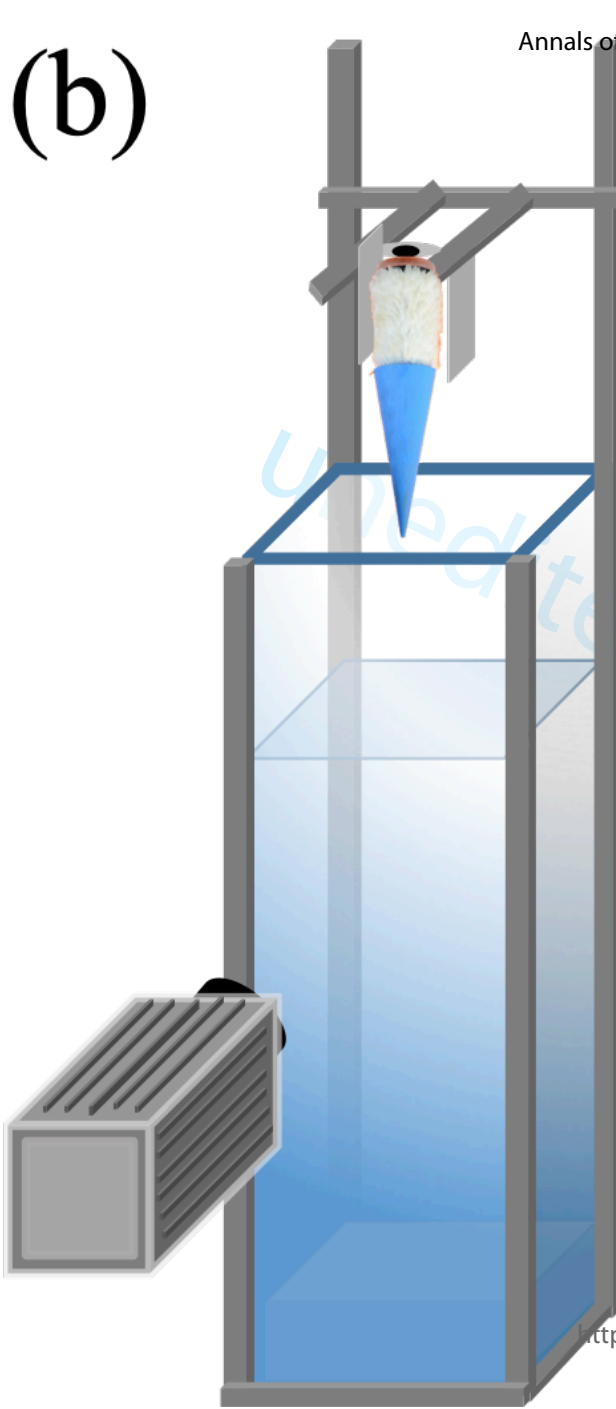
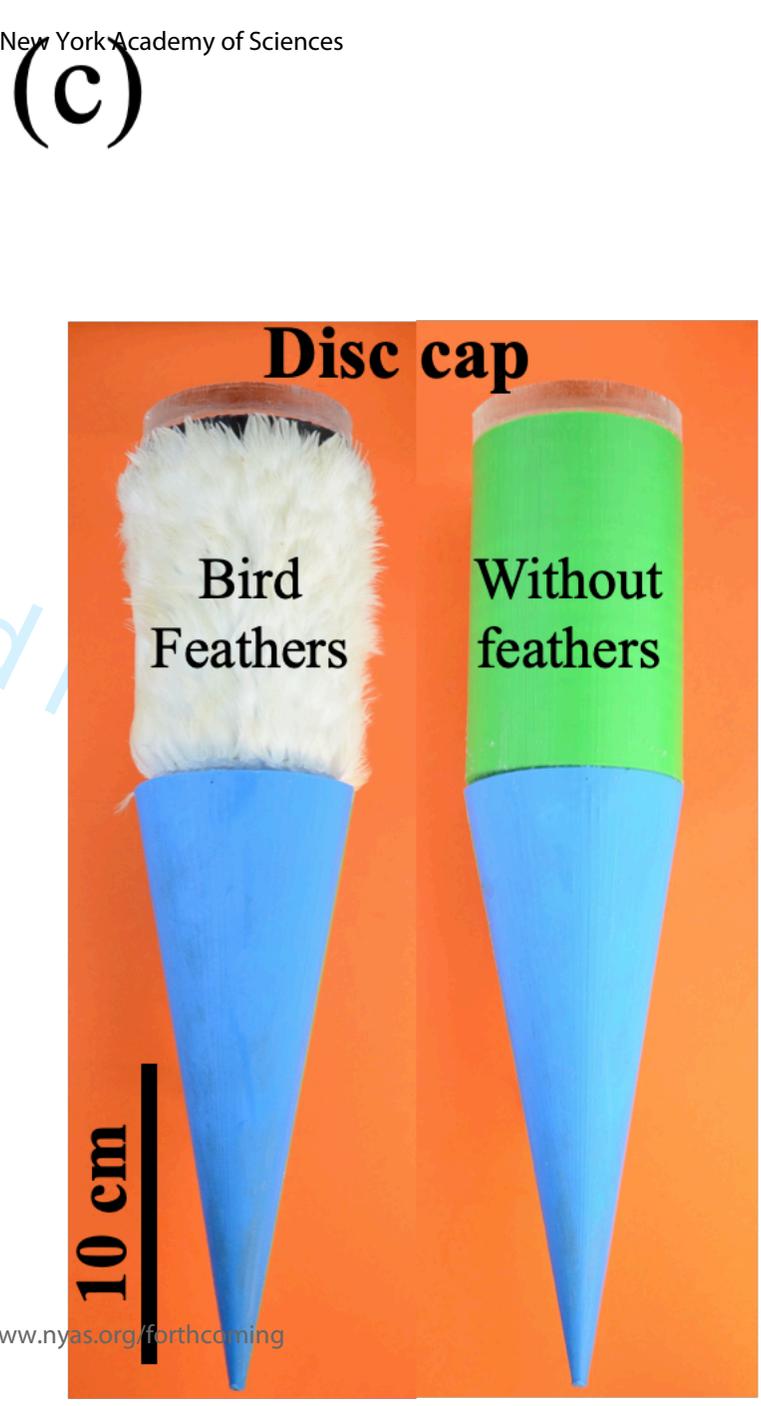
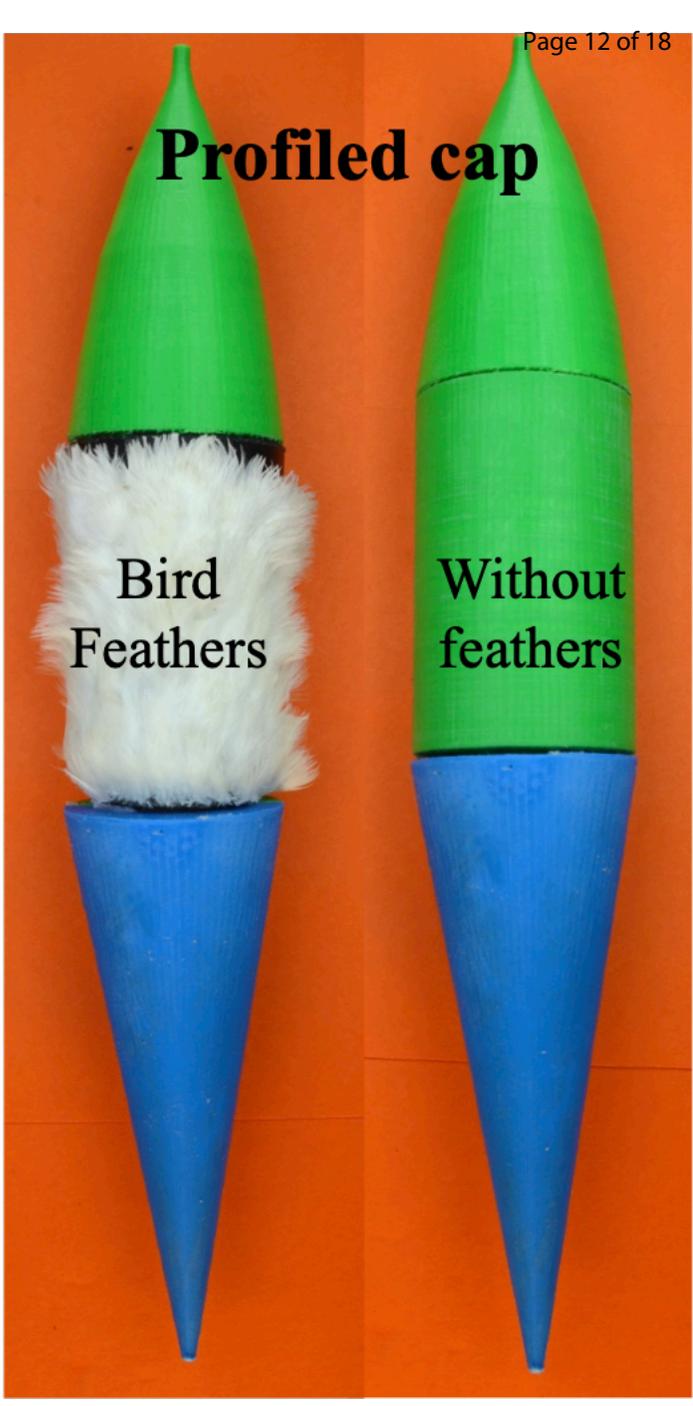

(a) (b) (c)

**Disc cap**

Bird Feathers

Without feathers

**Profiled cap**

Bird Feathers

Without feathers

10 cm

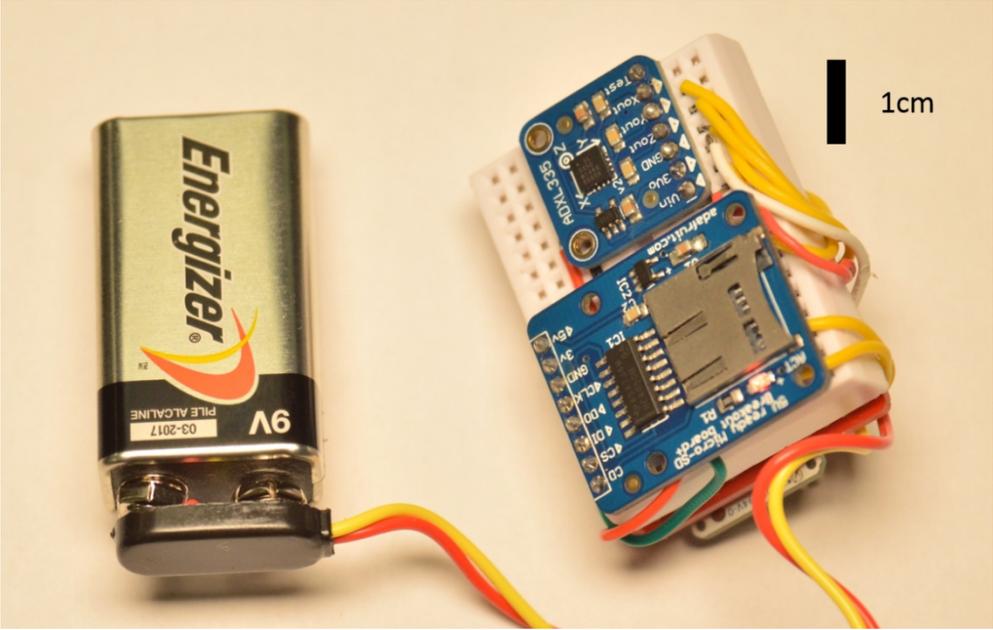

Photograph of the onboard accelerometer assembly, consisting of a microcontroller and an ADXL335 accelerometer module, alongside a standard 9V battery. The scale bar represents 1 cm.

503x318mm (59 x 59 DPI)

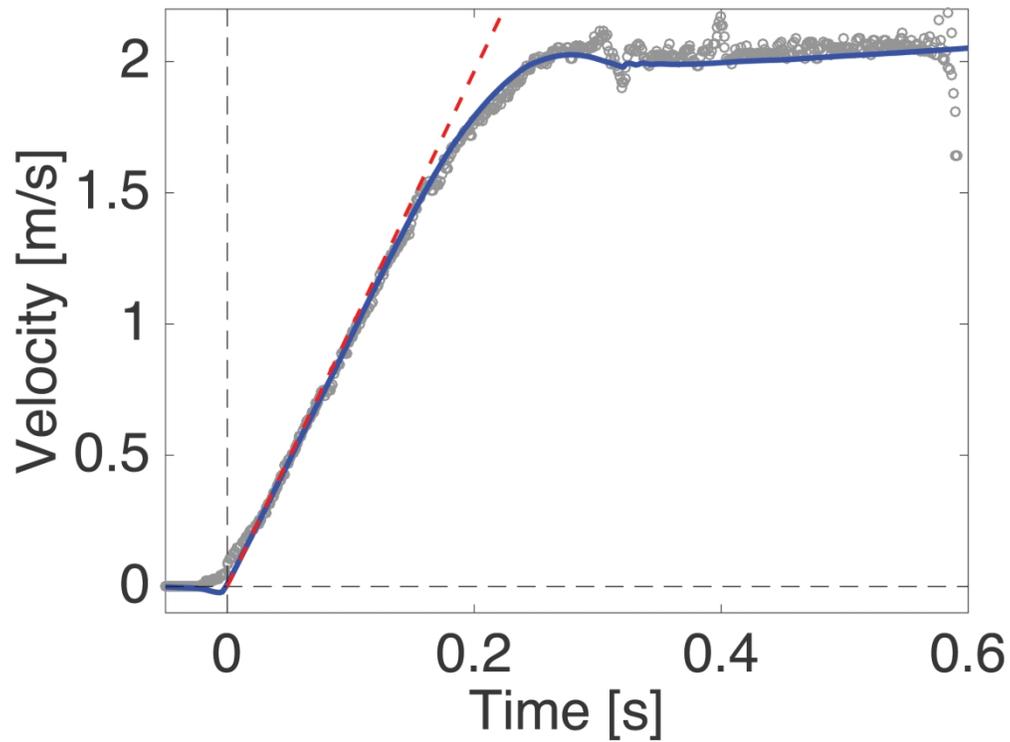

Velocity-time graph comparing measurements from the onboard accelerometer (solid line) and high-speed camera (circular dots). The red line shows a linear velocity increase due to gravity only.

487x359mm (118 x 118 DPI)

1
2
3
4
5
6
7
8
9
10
11
12
13
14
15
16
17
18
19
20
21
22
23
24
25
26
27
28
29
30
31
32
33
34
35
36
37
38
39
40
41
42
43
44
45
46
47
48
49
50
51
52
53
54
55
56
57
58
59
60

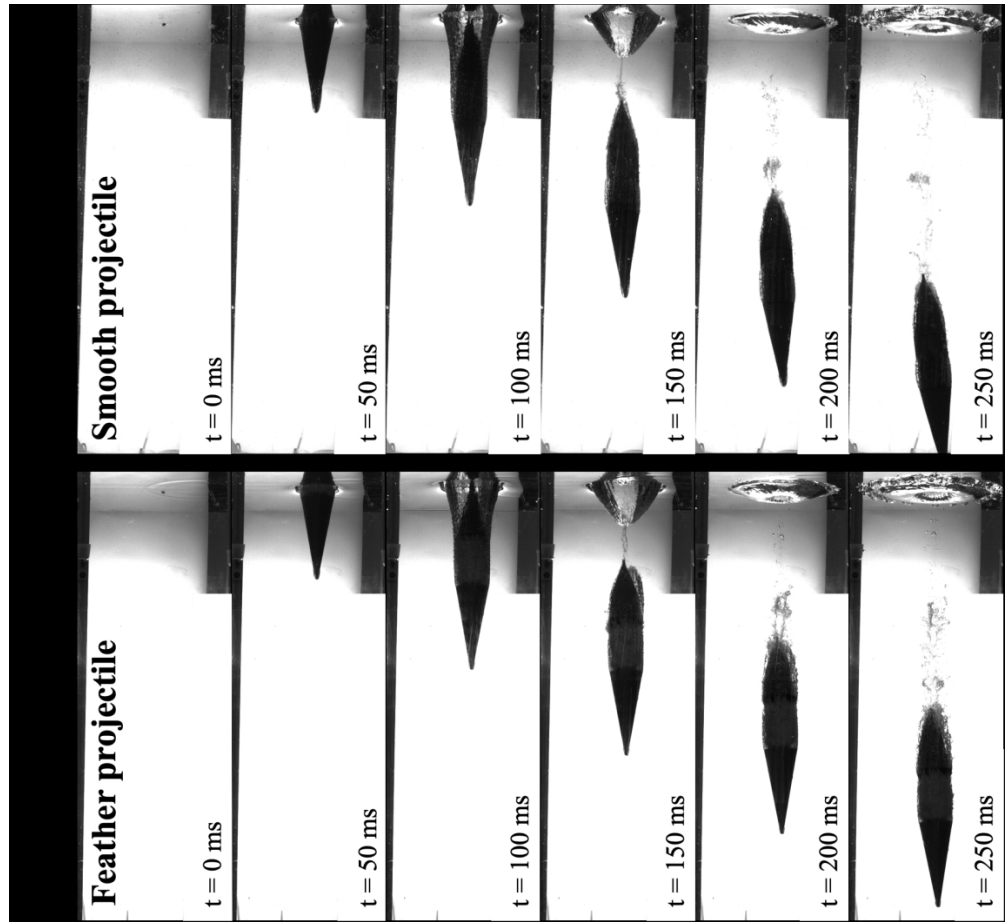

Sequential high-speed camera images documenting the water entry dynamics of two different projectile types. (a) The top row depicts the motion of a smooth projectile, while (b) the bottom row shows a feathered projectile. Each column represents a time step of 50 ms, starting from the moment of water entry at $t = 0$ ms and ending at $t = 250$ ms. The images capture the distinct interaction patterns of each projectile with the water, illustrating the variations in splash, cavitation, and water flow caused by the different surface textures.

321x295mm (144 x 144 DPI)

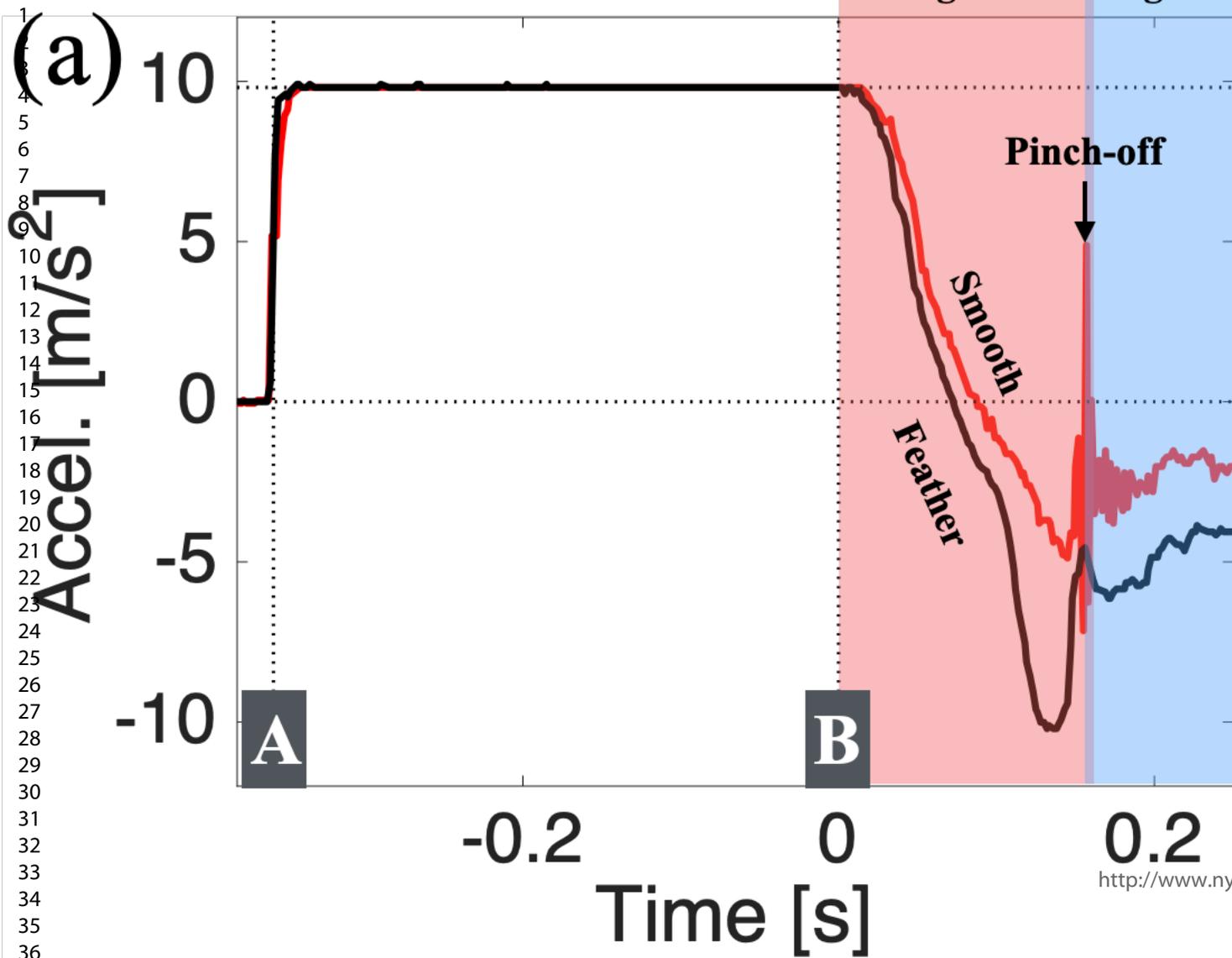
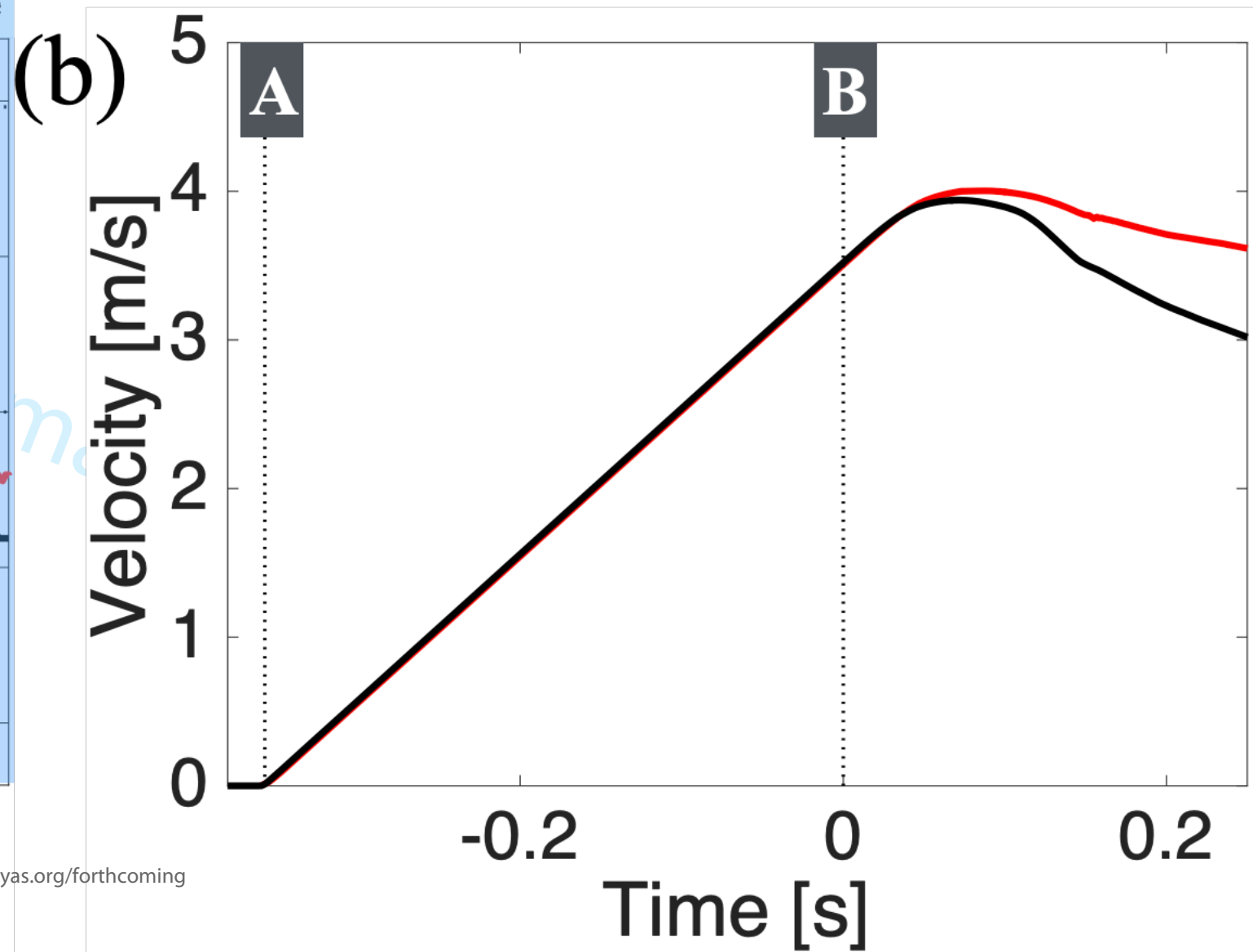

1
2
3
4
5
6
7
8
9
10
11
12
13
14
15
16
17
18
19
20
21
22
23
24
25
26
27
28
29
30
31
32
33
34

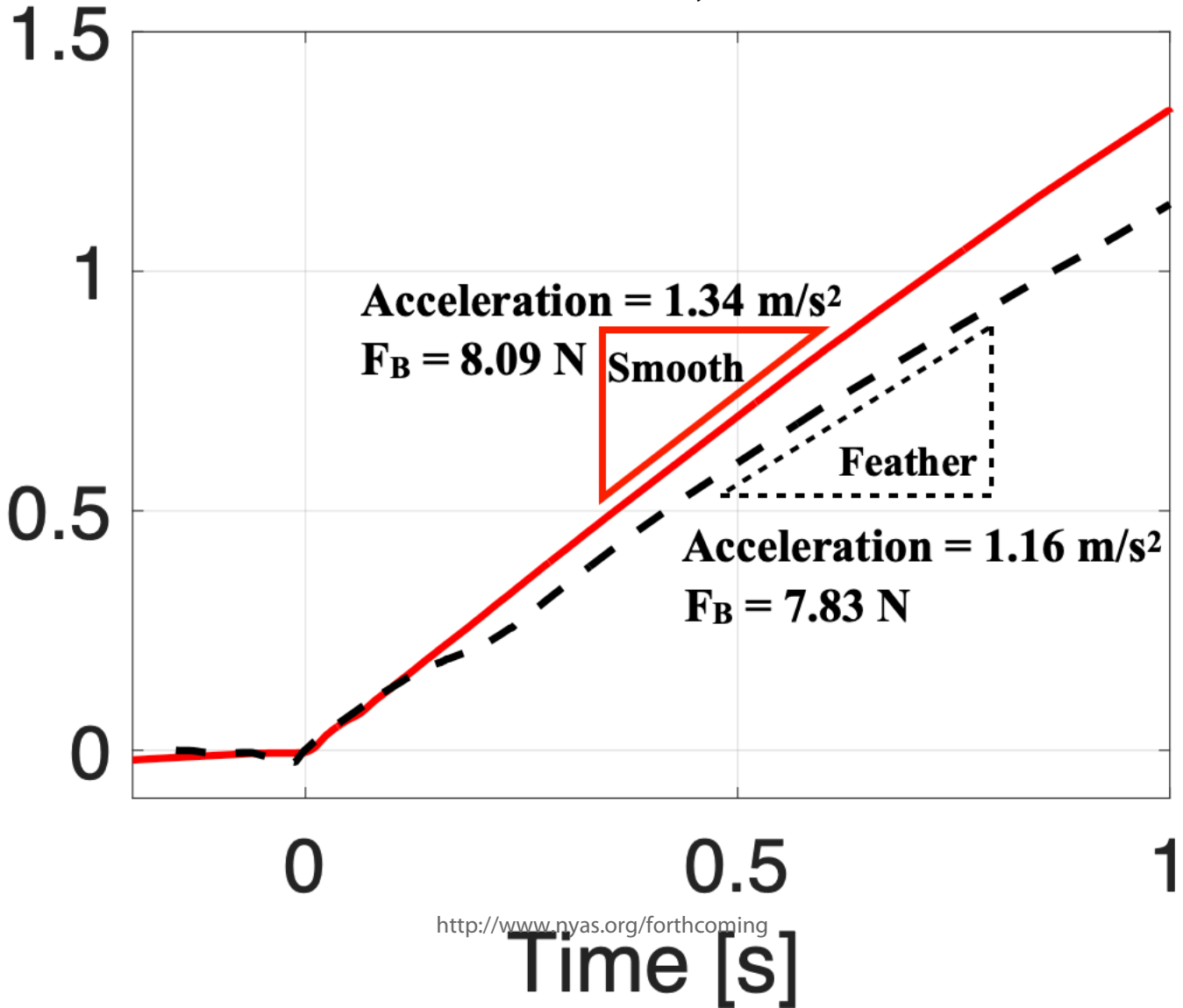

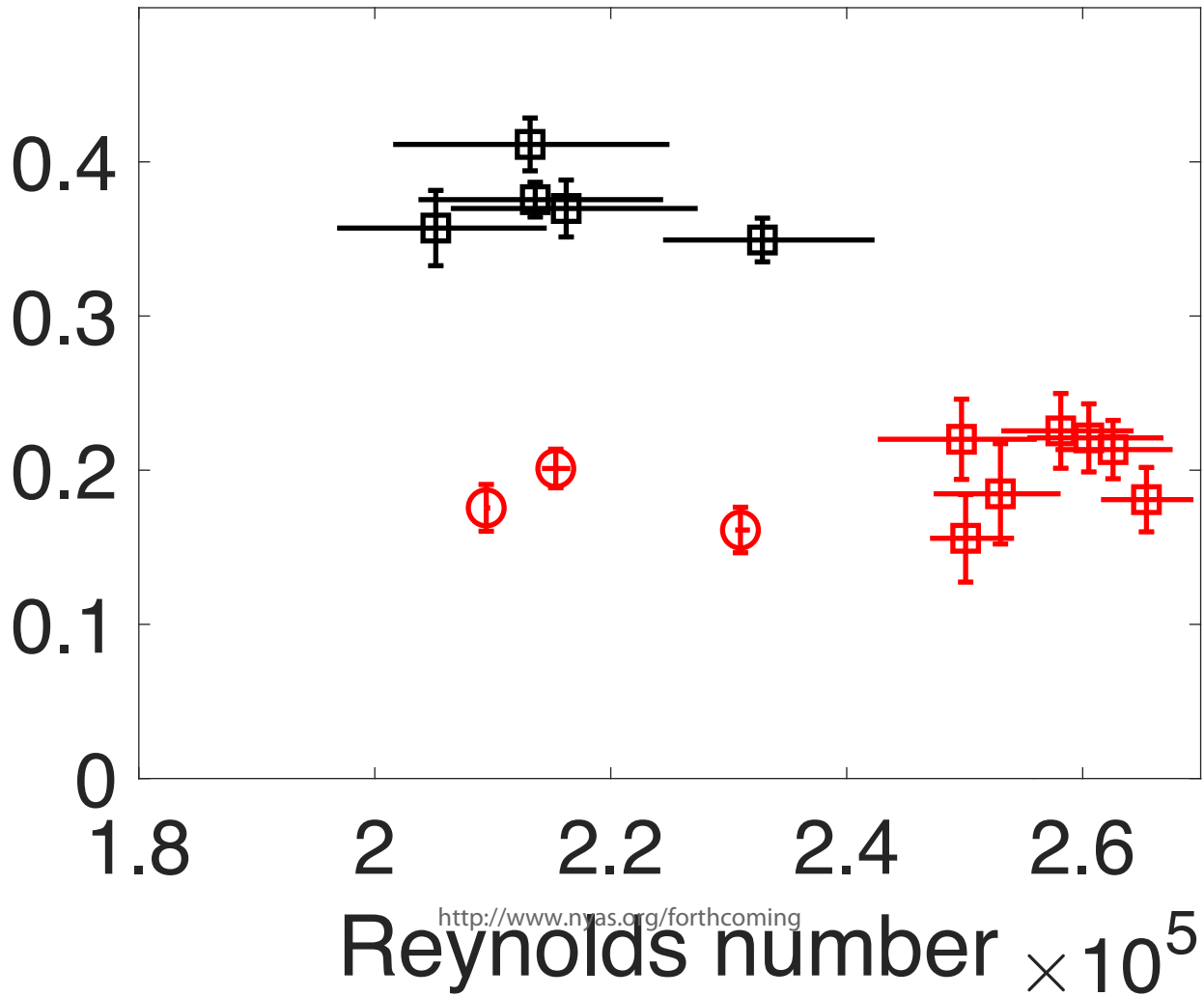